\documentclass[runningheads]{llncs}
\usepackage{amsmath,amssymb,tikz,enumitem}
\DeclareMathOperator{\C}{C} \newcommand{\Cev}{\C_\textnormal{ev}} \newcommand{\Codd}{\C_\textnormal{odd}}
\DeclareMathOperator{\Pev}{P_{\textnormal{ev}}} \DeclareMathOperator{\Podd}{P_{\textnormal{odd}}} \DeclareMathOperator{\Pc}{PC} 
\newcommand{\lar}{{\leftarrow}}

\title{Space-bounded online Kolmogorov complexity is additive
   \thanks{
     The work is supported by the HSE University Basic Research Program.}
   }
\author{Bruno Bauwens\inst{1}\orcidID{0000-0002-6138-0591} \and Maria Marchenko\inst{2}}
\institute{HSE University, Faculty of Computer Science, Moscow, 101000, Russia \email{brbauwens@gmail.com}\\ 
  \and University of Amsterdam, Amsterdam, Netherlands \email{maria.m.m.marchenko@gmail.com}
  }
\begin{document}
\maketitle
  
\begin{abstract}
  The {\em even online Kolmogorov complexity} of a string $x = x_1 x_2 \cdots x_{n}$ is the minimal length of a program that for all $i\le n/2$, 
  on input $x_1x_3 \cdots x_{2i-1}$ outputs $x_{2i}$. The {\em odd} complexity is defined similarly. 
  The sum of the odd and even complexities is called the {\em dialogue} complexity. 
  In~\cite{Bauwens2014asymmetry} it is proven that for all $n$, there exist $n$-bit $x$ for which the dialogue complexity exceeds the Kolmogorov complexity by $n\log \frac 4 3 + O(\log n)$. 
  Let $\C^s(x)$ denote the Kolmogorov complexity with space bound~$s$. 
  Here, we prove that the space-bounded dialogue complexity with bound $s + 6n + O(1)$ is at most $\C^{s}(x) + O(\log (sn))$, where $n=|x|$. 
\end{abstract}


\section{Introduction}

The Kolmogorov complexity of a bitstring $x$ is the minimal length of a program that prints~$x$. 
It was first used by Solomonoff to define an ideal algorithm for general sequence prediction~\cite{SolomonoffI,SolomonoffII,Solomonoff1960}. 
However, in real life situations, prediction often exploits side information. 
For example, one may aim to predict whether certain stocks go up or down relative to macro economic indicators. 
This corresponds to a complexity notion where one alternates between receiving information and generating output. 

The  {\em online} Kolmogorov complexity is a generalization of conditional complexity, introduced in~\cite{onlineComplexity}. 
Given two lists of strings $x_1, \ldots, x_k$ and $y_1, \ldots, y_k$, it is defined by the minimal length of a program that for all $j \le k$, 
on input $y_1, \ldots, y_j$ prints $x_j$. 
This notion has important applications in the foundations of reinforcement learning~\cite{Hutter2005universal,Hutter2024book}.\footnote{
  An agent and an environment exchange messages. The AIXI-agent defined in \cite{Hutter2005universal,Hutter2024book} models the environment using online a priori probability, 
  which is a probabilistic version of complexity. 
  The online coding theorem provides a close relation between the probabilistic and program complexity versions. 
  A space bounded variant of this theorem is essential to prove the main result below. 
}

A special case is the {\em even complexity} $\Cev(x)$, which is the task of predicting each next even bit in~$x$ given previous odd bits. 
The {\em odd} complexity $\Codd(x)$ is defined similarly. 
A time-bounded variant of this complexity is used in a recent combinatorial characterization of 2 cryptographic conjectures:
the existence of public key cryptography and one way functions~\cite{Ball2023Kolmogorov}.

Kolmogorov complexity satisfies an additivity property: the information in a pair $(x,y)$ of bitstrings equals the information in $x$ and the information in $y$ given~$x$, thus 
\begin{equation}\label{eq:symmetry}\tag{$*$}
 \C(x,y) = \C(x) + \C(y|x) + O(\log \C(x,y)). 
\end{equation}
Note that for Shannon entropy, the chain rule is exact $H(X,Y) = H(X) + H(Y|X)$. 
The additional non constant big-O term is (among other things) a consequence of the uncomputability of the complexity function~\cite{BauwensCompcomp,complexityOfComplexity}.  
Remarkably, such a result does not hold for online complexity: the sum of odd and even complexities can be much larger than the complexity of a string. 
Again, this is a consequence of the uncomputability of complexity. 

\begin{theorem}[\cite{Bauwens2014asymmetry,MuchnikOnline}]\label{th:BauwensOnline}
  There exists a sequence $\omega_1 \omega_2 \cdots$ such that for all $n$, 
  \[
    (\Cev + \Codd)(\omega_1 \cdots \omega_n) \ge n \log \tfrac 4 3 + \C(\omega_1 \cdots \omega_n). 
  \]
\end{theorem}

There is a close relation to Muchnik's paradox \cite{MuchnikOnline} in algorithmic randomness: 
if an infinite sequence is not random, one might neither blame the even nor the odd bits. 
Similar results for computable randomness and Hausdorff dimensions are given in~\cite{Barmpalias2021irreducibility,Barmpalias2022aspects}. 

In this paper, we study additivity properties for space bounded variants. 
Note that there is no hope to prove a time-bounded version of this result, because it would contradict the existence of one-way functions~\cite{Hirahara2023duality,Longpre1986PhD,Longpre1993symmetry}. 

The Kolmogorov complexity $\C^s(x|y)$ {\em with space-bound $s$,} 
is the minimal length of a program that prints~$x$ on input~$y$ and uses at most $s$ cells of memory~\cite{Bauwens2022Inequalities,Hartmanis1983generalizedKolm}, (see section~\ref{sec:def} for the full definition). 
It is well known that this complexity satisfies an additivity property. 
Note that the equality \eqref{eq:symmetry} can be viewed as two inequalities. 
With space-bounded complexity, the smaller side of each inequality needs a larger space-bound. 
The easy direction is
\[
  \C^{s+2|x|+O(1)}(x,y) \le \C^s(x) + \C^s(y|x) + O(\log \C^s(x)). 
\]
This is proven by simply running the program for $x$ and for $y$ given $x$. 
Also, the other inequality holds. 

\begin{theorem}[\cite{Bauwens2022Inequalities}]\label{th:spaceboundedSymmetry}
  For all strings $x,y$ and number $s$, 
  \[
    \C^{s'}(x) + \C^{s'}(y|x) \le \C^s(x,y) + O(\log \C^s(x,y)), 
  \]
  where $s' = s + O(|xy|)$.  
\end{theorem}

In the limit for large~$s$, this inequality reduces to additivity for unbounded complexity. 
Only the weaker result from \cite{Longpre1986PhD,Longpre1993symmetry} is well known: if $s' = 2s + O(|xy|+\log s)$, then $\C^{s'}(x) + \C^{s'}(y|x) \le \C^s(x,y) + O(\log s + \log \C^s(x,y))$. 
The proof follows the one for unbounded complexity.\footnote{
  There is a small caveat: the decompresser needs to iterate over pairs $(x,y)$ with $\C^s(x,y) < k$, and hence, iterate through all halting programs that use at most $s$ bits of memory. 
  To check whether a program halts, one needs an additional $s$ bits of memory. 
  The main idea of the proof of  theorem~\ref{th:spaceboundedSymmetry} is to do this more efficiently using Sipser's trick from~\cite{Sipser1980halting}: 
  given a halting configuration of a machine, consider all computations that terminate in this state, and organize all states in a tree. 
  Then do a DFS search starting in this halting configuration. 
} 
Because of the $O(\log s)$ in the right side, the result \eqref{eq:symmetry} for unbounded complexity is not a corollary. 

The main result of this paper is that additivity of online information does hold for space-bounded complexity. 
This is indeed possible, because space bounded complexity is computable (in small space), see~\cite[lemma 3]{Bauwens2022Inequalities} or proposition~\ref{prop:computeCs} below.  
Moreover, the gap $s'-s$ is very small. 

\begin{theorem}\label{th:main}
  For all $x$ and $s$, 
  \[
    \Cev^{s'}(x) + \Codd^{s'}(x) \le \C^s(x) + O\big(\log s + \log |x|\big),
  \]
  where $s' = s + 6|x| + O(1)$.
\end{theorem}

\noindent
{\em Remark.} It is not possible to drop the term $O(\log s)$, 
because otherwise, the limit for large $s$ would contradict theorem~\ref{th:BauwensOnline}.
Also, the proof gives a better precision of the big-O term at the right: $2d + 4\log d + O(1)$ where $d = \C^s(s,\C^s(x),|x|)$.\footnote{
  We believe it is possible to improve the $O(d)$ term to $O(\C_s(\C_s(s)))$ when using prefix or monotone complexity, and still 
  $s'-s \le O(|x|)$.  However, the proof seems significantly harder. 
  We hope to publish it in an extended version of this paper. 
}

\bigskip
\noindent
{\em Proof structure.} In the next section we define odd and even online semimeasures. 
We fix such semimeasures $\Pev$ and $\Podd$ that can be computed with $s + O(|x|)$ space and satisfy 
\begin{align*}
  \C^s(x) \ge \log \frac 1 {\Pev(x)} + \log \frac 1 {\Podd(x)}. 
\end{align*}
Let $k = \C^s(x)$. 
We prove a space-bounded version of the online coding theorem from~\cite{onlineComplexity}. This implies
\begin{align*}
  \Codd^s(x) &\le \log \frac 1 {\Podd(x)} + O(\C^s(s,|x|,k)) \\
  \Cev^s(x) &\le \log \frac 1 {\Pev(x)} + O(\C^s(s,|x|,k)). 
\end{align*}
The theorem follows by combining the 3 inequalities.

\section{Definitions}\label{sec:def}

For background in Kolmogorov complexity, see~\cite{LiVitanyiForthEdition,bookShenVereshchagin}. 
Typically, space-bounds in Kolmogorov complexity are defined and used up to arbitrary constant factors, 
see for example~\cite{Hartmanis1983generalizedKolm,Longpre1986PhD,Longpre1993symmetry}. 
In~\cite{Bauwens2022Inequalities}, a definition is given with additive precision $O(1)$. We use it here as well. 

We consider Turing machines that contain the following. 
\begin{itemize}
  \item For each input, there is a separate one-directional read-only tape with end markers. (Both for the program and each auxiliary string in the definition of complexity.)
  \item A one-directional write-only output tape. 
  \item Two binary stacks, with \textsc{push}, \textsc{pop} and \textsc{isempty} requests.
\end{itemize}

\noindent
{\em Remarks.}
\begin{itemize}[leftmargin=*, label=--]
  \item 
    A computation uses {\em space $s$} if the sum of the sizes of the 2 stacks is always at most~$s$. 
  \item 
    One may think of the 2 stacks as a standard work tape, where 1 stack represents the part at the left of the computation head and the other stack the right part. 
    Note that to make this analogy work, one needs to think about a binary work tape with some ``end marker'' that can be moved.\footnote{ 
      Popping an element from the stack corresponds to shifting all elements on one side of the tape, and this operation can be done if an endmarker is available. 
    }
    This definition allows for cleaner statements and proofs of space-bounded additivity (with $s' = s + O(|xy|)$ without extra $O(\log s)$ term). 
  \item 
    In proofs, one may use a more general type of machine, which is equipped with extra binary work tapes that have end markers. 
    Consider a machine with $k$ such tapes. Also, consider a computation on which at most $t_1, \ldots, t_k$ bits of these tapes are used, and the total stack size is at most~$s$. 
    This computation is simulated on a machine of the above type using space $s + \sum_{i\le k} t_i + 2\lceil \log t_i \rceil$. 
\end{itemize}

\newcommand{\argg}{x_1 \lar y_1; \ldots; x_k \lar y_k}
\begin{definition}\label{def:onlineComplexity}
  Let $x_1, y_1, \ldots, x_k, y_k$ be strings. 
  The {\em Kolmogorov complexity of tasks $x_1 \lar y_1$; \ldots; $x_k \lar y_k$ with space bound $s$} relative to machine $V$ is
  \[
    \C^s_V(\argg) = \min \big\{|p| : \forall j \mathop{\le} k \big[V(p,y_j) = x_j \textnormal{ using space } s\big]\big\}.
  \]
\end{definition}

\noindent
Note that even for $s = +\infty$, this definition is more general than in~\cite{onlineComplexity}, because only $y_j$ is used to produce~$x_j$, (not $y_1, \ldots, y_j$).  

\begin{proposition}
  There exists a machine $U$ with the following property. For each machine $V$ there exists a constant $c$ such that for all $s,k$ and strings $x_1,y_1, \ldots, x_k, y_k$, 
  \[
    \C_U^{s+c}(\argg) \le \C_V^s(\argg) + c. 
  \]
\end{proposition}

\noindent
The proof follows the one of \cite[proposition 1]{Bauwens2022Inequalities}, 
where a universal machine is constructed in which a self-delimiting description of the simulated machine $V$ is prepended to its program. 
This description is stored on an auxiliary tape. The constant $c$ is the extra space needed by the auxiliary tape (in the simulation on the stacks). 

We fix such a machine $U$ and drop subscripts $U$ in all complexities. 

\begin{definition}
  The {\em conditional} complexity of string $x$ given string $y$ is $\C^s(x|y) = \C^s(x \lar y)$. 
  If $y$ is empty, we write $\C^s(x)$. 
  We use the prefix code $\bar y = 0y_1 \cdots 0y_{|y|}1$ and encode a tuple of strings $(t_1, \ldots, t_e)$ as $\bar t_1 \bar t_2 \cdots \bar t_{e-1} t_e$.
  The conditional complexity with tuples is defined using this encoding, thus $\C^s(y,z|x) = \C^s(\bar y z|x)$. 

  The {\em even} complexity of $x$ is 
  \begin{align*}
    \Cev^{s}(x) 
    &= \min \big\{|p|: \forall j \le |x|/2\big[U(p, x_1 x_3 \cdots x_{2j-1}) = x_{2j} \textnormal{ using space } s\big]\big\}\\
    &= \C^s(x_2 \lar x_1; \,x_4 \lar x_1x_3; \,\ldots).
  \end{align*}
  The  {\em odd} complexity of $x$ is defined similarly.  
  For unbounded complexity, thus $s = +\infty$, we drop $s$, for example $\C^{\infty}(x|y) = \C(x|y)$. 
\end{definition}

\medskip
\noindent
{\em Remark.} The  {\em decision} variant of the Kolmogorov complexity of $x$, is the minimal length of a program that prints 
some string or infinite sequence that has $x$ as prefix. It would be fair to compare $\Cev(x) + \Codd(x)$ to this variant, 
since the odd and even complexity do not require to indicate where $x$ stops. 
However, the decision and plain complexities differ by at most $O(\log \C(|x|))$, 
and this is less than the precision $O(\log (s|x|))$ of the main result. Hence, there is no reason to deviate from the more standard notions.

\section{Even and odd online measures}

\begin{definition}
  Let $P$ be a mapping from strings to the real interval $[0,1]$. 
  \begin{itemize}
    \item 
      $P$ is a {\em semimeasure} if for all~$x$, 
      \[
	P(x) \ge P(x0) + P(x1).
      \]
    \item 
      $P$ is an {\em even semimeasure} if for all~$x$, 
      \begin{eqnarray*}
	\textnormal{if $|x|$ is even } & \Rightarrow & P(x) = P(x0) = P(x1) \\
	  \textnormal{if $|x|$ is odd }  & \Rightarrow & P(x) \ge P(x0) + P(x1). 
      \end{eqnarray*}
    \item 
      $P$ is an {\em odd semimeasure} if for all~$x$, 
      \begin{eqnarray*}
	\textnormal{if $|x|$ is odd } & \Rightarrow & P(x) = P(x0) = P(x1) \\
	  \textnormal{if $|x|$ is even }  & \Rightarrow & P(x) \ge P(x0) + P(x1). 
      \end{eqnarray*}
  \end{itemize}
\end{definition}

\noindent
An online coding theorem for lower-semicomputable even semimeasures was proven in~\cite{onlineComplexity}. 
However, it has a logarithmic term in the right side. 
Below, we prove a variant for computable even semimeasures without this term. 

\begin{proposition}\label{prop:onlineCoding}
 For each computable even semimeasure $\Pev$ there exists a constant $c$ such that for all~$x$, 
  \[
    \Cev(x) \le \log \frac 1 {\Pev(x)} + c. 
  \]
\end{proposition}

\noindent
The goal of this section is to prove a version of this theorem with a space-bound. 
We use the following. 

\begin{definition}
  The complexity $\C^{\tilde s}(F)$ of $F: \{0,1\}^* \rightarrow \mathbb R$ with space bound $\tilde s: \mathbb N \times \mathbb N \rightarrow \mathbb N$ is the minimal length of a program on $U$ that on input $(k,x)$ uses space $\tilde s(k,|x|)$ and outputs an integer $N$ such that $|F(x)-N2^{-k}| \le 2^{-k}$, where $k$ is given in binary.  
  If no such program exists, then $\C^{\tilde s}(F) = +\infty$. 
\end{definition}

\begin{proposition}\label{prop:onlineCodingSpaceBounded}
  Assume $\tilde s$ is non-decreasing in both arguments. 
 There exists a $c$ such that for every even semimeasure $\Pev$ and string $x$, 
  \[
    \Cev^{s'}(x) \le \log \frac 1 {\Pev(x)} + O(\C^{\tilde s}(\Pev)),  
  \]
  where $s' = \tilde{s}(\big\lceil\log \frac {8|x|}{\Pev(x)}\big\rceil,|x|) + |x| + O(\log |x|)$. 
\end{proposition}

\noindent
A similar result for~$\Podd$ holds with a similar proof. 

\begin{proof}[of proposition~\ref{prop:onlineCoding}] 
  We first explain that any real even semimeasure can be transformed to a rational one that is at most a constant factor smaller. 
  Indeed, we may assume that $\Pev(x) \ge 2^{-\lfloor |x|/2\rfloor - 1}$, otherwise, we replace it by $2^{-\lfloor |x|/2\rfloor - 1} + \tfrac 1 2 \Pev(x)$. 
  Then, it is sufficient to round the value down with precision $2^{-|x|}$ for even $|x|$, and rescale recursively. 
  This decreases the semimeasure by at most a factor
  $
    \prod_{i=1}^\infty (1 - 2^{-i}) > 0. 
    $

  We now assume that $\Pev$ is rational. 
  Let the cumulative semimeasure $\Pc(y)$ represent the measure of all $z$ with $|z|=|y|$ such that $0.z < 0.y$, 
  more precisely, let $\Pc(\textnormal{empty str}) = 0$ and
  \[
    \Pc(yb) = 
	\begin{cases}
	  \Pc(y) & \text{ if } |y| \textnormal{ is even or } b=0, \\
	  \Pc(y) + \Pev(y0) & \textnormal{ otherwise, thus $|y|$ is odd and $b=1$.}
	\end{cases}
  \]
  The decompresser depends on this function. 
  For notational convenience, consider a decompresser in the definition of even complexity that also takes the previous answers as input. 
  For example, in $\Cev(y)$, on input $y_1y_2y_3y_4y_5$ the program outputs $y_6$, (instead of input $y_1y_3y_5$). 

  \medskip
  \noindent
  {\em Machine $V$ on input $p$ and $y$:
  if $2^{-|p|}p < \Pc(y1)$ output 0, otherwise output 1.} 

  \medskip
  \noindent
  Recall that $\Pev$ is a computable rational function. 
  Hence, also $\Pc$ is a computable rational function, 
  because $\Pc(y)$ equals the sum of at most $|y|/2$ values~$\Pev(\cdot)$. 
  Thus $V$ computes 2 rational numbers and outputs the result of their comparison. 

  
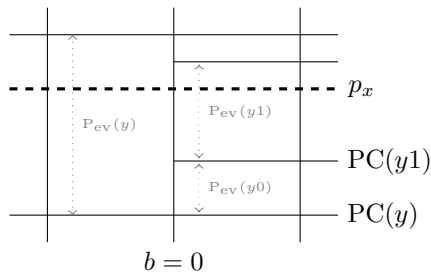
\begin{center}
    \begin{tikzpicture}[xscale=1.4,scale=1.2]
      \draw (0,-.3) -- +(0,2.6);
      \draw (1,-.3)  node[below] {$b=0$}-- +(0,2.6);
      \draw (2,-.3) -- +(0,2.6);
      \draw (-.3,0) -- +(2.6,0) node[right] {$\Pc(y)$};
      \draw (-.3,2) -- +(2.6,0);
      \draw (1,0.6) -- +(1.3,0) node[right] {$\Pc(y1)$};
      \draw (1,1.7) -- +(1.3,0);
      \draw[dotted,<->,gray] (0.2,0) -- node[fill=white,right] {\tiny $\Pev(y)$} +(0,2);
      \draw[dotted,<->,gray] (1.2,0.03) -- node[fill=white,right] {\tiny $\Pev(y0)$} +(0,0.54);
      \draw[dotted,<->,gray] (1.2,0.63) -- node[fill=white,right] {\tiny $\Pev(y1)$} +(0,1.04);
      \draw[dashed, very thick] (-.3,1.4) -- +(2.6,0) node[right] {$p_x$};
    \end{tikzpicture}
  \end{center}
  
  \noindent
  We verify that if $y$ is a strict prefix of $x$ with odd length and if 
  \[
    p2^{-|p|} \in \big[\Pc(x), \Pc(x) + \Pev(x)\big), 
    \]
  then $V(p,y) = x_{|y|+1}$. This is quite logical, but we give the details anyway. Fix a prefix $z$ of $x$, and let $b = x_{|z|+1}$. If $b=0$, then 
  \[
    \big[\Pc(zb), \Pc(zb) + \Pev(zb)\big) \;\subseteq\; \big[\Pc(z), \Pc(z) + \Pev(z)\big). 
  \]
  This also holds for $b=1$, by expanding the definition of~$\Pc(z1)$. It also holds for $|z|$ of even length. 
  For $b = x_{|y|+1}$, it holds that
  \[
    p2^{-|p|} \;\in\; \big[\Pc(x), \Pc(x) + \Pev(x)\big) \;\subseteq\; \big[\Pc(yb), \Pc(yb) + \Pev(yb)\big). 
  \]
  We verify that $b = 0$ if and only if $2^{-|p|}p < \Pc(y1)$: 
  \\-- If $b = 0$, then $p2^{-|p|} < \Pc(y0) + \Pev(y1) = \Pc(y1)$. 
  \\-- If $b = 1$, then $p2^{-|p|} \ge \Pc(y1)$. 
  \\This implies $V(p,y) = b$. 

  \medskip
  To finish the proof, consider any $x$ with $\Pev(x)>0$. Let $k_x = \lceil \log \frac 1 {\Pev(x)} \rceil$ and let
  \[
    p_x = \lceil 2^{k_x} \cdot \Pc(x) \rceil
  \]
  be a number in binary with $k_x$ bits (if needed, add leading zeros). 
  Note that $2^{-k_x}p_x$ lies in the interval $\big[\Pc(x), \Pc(x) + \Pev(x)\big)$, because $\Pev(x) \ge 2^{-k_x}$. 
  Thus, 
  \[
    \C_{\textnormal{ev},V}(x) \le |p_x| = k_x = \Big\lceil \log \frac 1 {\Pev(x)} \Big\rceil. 
  \]
  \qed
\end{proof}

\begin{proof}[of proposition~\ref{prop:onlineCodingSpaceBounded}]
  Now $\Pev$ is a real valued function and computable with space bound $\tilde s$. 
  The idea is to use the program $p'_x = p_x01$, which represents a number in the interval $\big[\Pc(x), \Pc(x)+\Pev(x)\big)$ that is separated from the borders by at least $2^{-|p_x|-2}$.  
  We use the same algorithm for $V$, but $\Pc(y1)$ is calculated with precision $2^{-|p_x|-2}$. 

  For this, all of the at most $|y|/2$ used values of $\Pev$ are computed with $k_x + 2 + \lceil \log |y|\rceil$ bits precision, where $k_x = \lceil \log \Pev(x) \rceil$. 
  This requires space  
  \[
    \tilde s(k_x + 2 + \lceil \log |x|\rceil, |x|) + k_x + \log |x| + O(1). 
  \]
  \qed
\end{proof}

\section{Proof of theorem~\ref{th:main}}

In~\cite[lemma 3]{Bauwens2022Inequalities} it is shown that on input $x$ and $s$, one can compute $\C^s(x)$ in space $s + O(|x|)$. 
With a similar proof we obtain the following. 

\begin{proposition}\label{prop:computeCs}
  There exists a machine that on input $x,k,s$ decides whether $\C^s(x) < k$ using space $s+2|x|+O(\log |x|)$.  
\end{proposition}

\noindent
The proof of  theorem~\ref{th:main} relies on defining an odd and an even semimeasure. They are derived from the following (standard) semimeasure. Fix $s,k$ and $n$. Let
\[
  Q(y) = \frac {\big|\big\{z : |yz| = n \textnormal{ and } \C^s(yz) \mathop< k \big\}\big|} {2^{k}} .
\]
{\em Remarks.} 
\\-- At most $2^k-1$ strings have complexity less than $k$, thus $Q(\text{empty}) < 1$. 
\\-- $Q$ is indeed a semimeasure with $Q(x) = 0$ if $|x| > n$. 
\\-- If $\C^s(x) < k$ and $|x|=n$, then $-\log Q(x) \le k$. 
\\-- The space needed to evaluate $Q(y)$ exactly is 
\[
  (n + O(\log n)) + (s + 2n + O(\log n)), 
\]
where the first term, is for storing the counter $z$ (in the definition of $Q$), 
and the second is the bound of proposition~\ref{prop:computeCs} for checking~$\C^s(yz)<k$.  

For $x$ of even length~$\ell$, the odd and even semimeasures are 
\begin{align*}
  \Podd(x) &= Q(x_1) \cdot \frac {Q(x_1x_2x_3)}{Q(x_1x_2)} \cdot \ldots \cdot \frac {Q(x_1 \cdots x_{\ell-2}x_{\ell-1})}{Q(x_1 \cdots x_{\ell-2})} \\
  \Pev(x) &= \frac {Q(x_1x_2)} {Q(x_1)} \cdot \frac {Q(x_1x_2x_3x_4)}{Q(x_1x_2x_3)} \cdot \ldots \cdot \frac {Q(x_1 \cdots x_{\ell-1}x_{\ell})}{Q(x_1 \cdots x_{\ell-1})}, 
\end{align*}
and these semimeasures are then uniquely defined on strings of odd lengths.  

\medskip
\noindent
{\em Remarks.} 
\\-- $Q(x) = \Podd(x)\Pev(x)$.
\\-- For $\tilde s(k',n') = s + 3n' + k' + O(\log n')$, we prove that
\[
  \C^{\tilde s}(\Podd) = \C^s(s,n,k) + O(1).  
\]
Indeed, if $|y| > n$, then $\Podd(y) = 0$. Otherwise, $\Podd$ is the product of $|y| = n'$ numbers, and each is computed and stored with precision $k' + \lceil\log n' \rceil$. 

\medskip
\noindent
Now we finish the proof of the main theorem. For some $x$ and $s$, consider $Q$ with parameters $s = s,\, k = \C^s(x) + 1$ and $n = |x|$. 
We may assume $k \le n$, otherwise the theorem is trivial.  For $s' = \tilde s(n + 3 + \lceil \log n \rceil,n) + n + O(\log n)$, 
\begin{align*}
  \C^s(x) + 1 & \ge - \log Q(x) \\ & = - \log \Big(\Podd(x) \, \Pev(x)\Big) \\&\ge \Cev^{s'}(x) + \Codd^{s'} - O(\C^s(s,n,k)).
\end{align*}
Since $s' \le s + 6n + O(1)$, this finishes the proof of theorem~\ref{th:main}.

\subsection*{Acknowledgements}

The authors are grateful to an anonymous referee for suggesting a better title and careful comments.

\bibliographystyle{plain}
\bibliography{kolmogorov}

\end{document}